# Uncertainty Discounting in Deterministic Black Box Price Predictions for Energy Arbitrage

Arnab Bhattacharjee
Power, Energy and Control Discipline
School of Electrical Engineering and Computer Science
The University of Queensland
St.Lucia, Brisbane, Australia
a.bhattacharjee@uq.edu.au

*Abstract*— **This study examines the economic impact of post-hoc uncertainty discounting in predictive energy management, specifically in battery energy arbitrage. A 2.2 MWh/1.1 MW Tesla battery, emulating operations at the University of Queensland's St. Lucia campus, is used as a test system. Traditionally, Model Predictive Control (MPC) frameworks rely on deterministic spot price forecasts from the Australian Energy Market Operator (AEMO) to optimize battery scheduling. However, these forecasts lack uncertainty awareness, making arbitrage strategies vulnerable to extreme price volatility. To address this, we propose simple heuristic uncertainty discounting methods, which require no access to the predictive model's architecture or inputs. By integrating these strategies into existing MPC frameworks, we demonstrate a more than 20% improvement in economic returns under identical operational constraints. This approach enhances decision-making in energy arbitrage while remaining practical, scalable, and independent of specific forecasting models.**

## I. INTRODUCTION

Uncertainty awareness is crucial in applications such as energy arbitrage and predictive energy management, particularly in systems with high renewable energy penetration. The intermittency of renewables introduces volatility in the energy-supply balance, leading to price fluctuations. Queensland experiences some of the highest energy price volatility in Australia, driven by widespread rooftop solar adoption and increasing extreme weather events [1]. In 2024, Queensland recorded 185 high-price volatility events where electricity prices exceeded $500/MWh (Figure 1). Such extreme fluctuations make predictive energy management tasks like energy arbitrage challenging, as these events are inherently difficult to anticipate. Such extreme non-stationary and volatile behaviours cannot be effectively captured by standard price forecasting tools, which are widely based on stationarity assumptions, limiting their ability to anticipate rapid fluctuations.

These factors require incorporating uncertainty-awareness in predictive energy management applications[2][3]. However, a significant hurdle in uncertainty-aware decision-making is the dependence on third-party price forecasting tools [4], over which decision-makers have no administrative control. Most third-party tools provide deterministic price forecasts without accompanying uncertainty metrics, making it difficult to factor confidence levels into downstream predictive management and control strategies [5]. While extensive research exists on post-hoc uncertainty quantification for deterministic predictive models, most methods expect either white box access to the predictive model's architecture and parameters [5][6] or complete input-output data access through either querying the model or access to the training dataset [6-8]. This significantly limits their potential in more practical and industrial applications where the decision-makers lack access to either input features or the underlying predictive model, relying solely on forecast outputs from black-box tools. Some works propose completely new forecasting tools that come with inherent uncertainty measures [8-10] but this work doesn't consider them as it is often practically limiting - due to unavailability of data, computational resources or expertise - for decision makers to build their own forecasting tools.

To address these challenges, this study proposes simple, computationally efficient yet effective uncertainty-discounting strategies that are designed based on widely applicable heuristic principles and can operate independently of predictive model architecture or input data. These methods serve as plug-and-play heuristics, enabling decision-makers to integrate uncertainty awareness into predictive management tasks when using third-party deterministic forecasts. A case study of energy arbitrage using a 2.2 MWh/1.1 MW Tesla battery, emulating the operations of one situated at the University of Queensland's St. Lucia campus, is considered for demonstration. Traditionally, the battery schedule is optimized using a Model Predictive Control (MPC) framework that leverages AEMO's pre-dispatch spot market price forecasts for up to 40 hours into the future without considerations for uncertainty in the forecasts. It is shown that by incorporating simple heuristic uncertainty discounting strategies in the existing MPC framework, the economic benefits accrued from energy arbitrage under the same operational constraints can be increased by more than 20%.

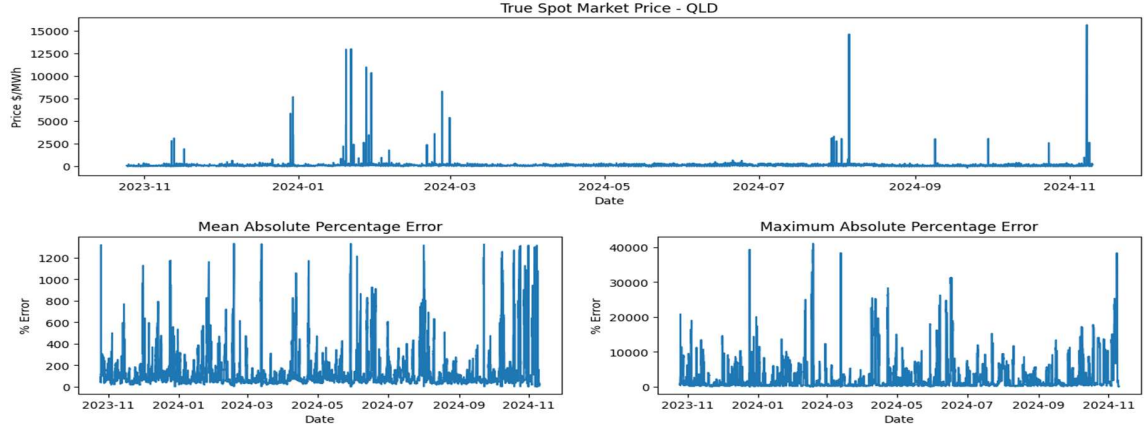

**Figure 1**: (Top)The spot market dispatch prices from Nov 2023 – Nov 2024 in QLD.(Bottom Left) Mean Absolute Percentage errors in AEMO's pre-dispatch forecasts. (Bottom Right) Maximum Absolute Percentage Errors in pre-dispatch forecasts[14]

The remainder of the paper is organized as follows: Section II presents the necessary background to understand AEMO's pre-dispatch price forecasting approach and outlines the standard MPC framework for optimized battery arbitrage, Section III details the proposed uncertainty discounting techniques and the experimental details, Section IV presents results and discussions, and Section V concludes the study with future research directions.

## II. Background

### A. AEMO's Pre-Dispatch Price Forecasting Mechanism

The Australian Energy Market Operator (AEMO) employs a sophisticated National Electricity Market Dispatch Engine (NEMDE) to generate pre-dispatch spot market price forecasts, providing market participants with short-term price and demand projections for informed decision-making in energy trading and operational planning [4].

AEMO's pre-dispatch forecasting operates on two primary time scales:

1. 5-Minute Pre-Dispatch – Updated every 5 minutes, offering a one-hour look-ahead for short-term price and demand forecasts.

2. 30-Minute Pre-Dispatch – Updated half-hourly, extending forecasts until the end of the next market day (up to 40 hours in half hour intervals).

The NEMDE engine optimizes dispatch decisions by solving a large-scale mixed-integer linear programming (MILP) problem, incorporating Generator bids and offers, Network constraints and interconnector flows, Demand forecasts and Ancillary service requirements. AEMO also integrates numerical weather prediction (NWP) models into its forecasting process to improve renewable generation forecasts, particularly for solar and wind power. These models enhance the accuracy of intermittent generation forecasts, ensuring that weather-driven fluctuations are accounted for in dispatch planning [11].

Despite the sophistication of AEMO's forecasting system, its pre-dispatch price forecasts remain deterministic, meaning they provide single-point predictions without explicit uncertainty quantification. While AEMO's dispatch process accounts for forecast variability, the published pre-dispatch price forecasts do not include confidence intervals or probabilistic uncertainty metrics. Figure 1 depicts the mean and maximum absolute percentage errors in the pre-dispatch price forecasts between Nov'23 – Nov'24. The percentage errors are staggering, making it challenging to use these forecasts in predictive energy management tasks without making considerations for uncertainty.

### B. Model Predictive Control for Battery Arbitrage

This study emulates The University of Queensland's Tesla Powerpack battery system (1.1 MW / 2.2 MWh) at the St. Lucia campus using publicly available information. The battery leverages a Model Predictive Control (MPC) framework to optimize charging and discharging schedules. The MPC framework integrates AEMO's pre-dispatch spot price forecasts to make real-time trading decisions, ensuring the battery charges when prices are low and discharges when prices are high. Further information on the battery operations can be found in [12][13].

In this study a simple MPC problem is formulated to emulate the optimized battery operations using AEMO's pre-dispatch prices:

$$\max_{P_k} \quad c_k^T P_k \qquad \text{- (1)}$$

s. t.

$$E_k^n = E_k^0 - \left(\frac{\eta \Delta_t}{E_{nom}}\right) P_k^n \quad \forall\, n \in [1, \dots, T_k] \qquad \text{- (2)}$$

$$E_k^0 = E_{k-1}^1 \quad\quad\quad\quad\quad\quad\quad\quad\quad\quad\quad\quad (3)$$
$$P_l \leq P_k^n \leq P_u \quad\quad \forall\, n \in [1, \dots, T_k] \quad\quad (4)$$
$$E_l \leq E_k^n \leq E_u \quad\quad \forall\, n \in [1, \dots, T_k] \quad\quad (5)$$
$$\mathbf{1}^T |\mathbf{P_k}|\Delta_t \leq \frac{T_k}{48} E_{nom} \quad\quad\quad\quad\quad\quad (6)$$

where, $T_k$ is the length of the prediction horizon at the present half hourly instant $k$. The value of $T_k$ varies between 32 and 80 depending on the time of the day, $k$. $\Delta_t = t_{[k,k+1)}$ represents the length of the interval between two consecutive instances, i.e., half hour. $E_{nom}$ is the nominal capacity of the battery. $\mathbf{P_k} = [P_k^1, P_k^2, \dots, P_k^n, \dots, P_k^{T_k}]^T$ is the power schedule of the battery over the entire horizon of the MPC at instant $k$ and $\mathbf{E_k} = [E_k^1, E_k^2, \dots, E_k^n, \dots, E_k^{T_k}]^T$ is the state of charge/energy remaining in the battery over the forecast horizon of $T_k$ steps at instant $k$ where $n$ represents the lead time. $\mathbf{c_k} = [c_k^1, c_k^2, \dots, c_k^{T_s}]$ are the pre-dispatch spot price forecasts over $T_k$ timesteps at instant $k$. $P_l, P_u, E_l, E_u$ are the lower and upper limits of the battery power output and SOC respectively.

At every half hourly instant, $k$, the MPC (1) is solved to determine the optimal power schedules for upto $T_k$ instants into the future. However, only the first element of the optimal power schedule vector, i.e., $P_k^1$, is implemented for the current half hour interval while the remaining are discarded. This process is repeated every half hour with newly available pre-dispatch price forecasts. The objective function maximizes the spread or the expected benefits accrued over the entire prediction horizon. Constraint (2) determines the SOC or the remaining energy in the battery, Constraint (3) ensures that the initial SOC of the battery at instant $k$ is equal to the remaining energy in the battery after the previous instant, $k-1's$ first value in the optimal power schedule has been executed. Constraints (4) and (5) represent the power and SOC limits and Constraint (6) ensures that the battery does not undergo more than one full charge throughput in a day. This is done to minimize the degradation in the battery due to over-cycling.

It is straightforward that the absence of uncertainty-aware forecasting in AEMO's pre-dispatch system means that the energy arbitrage strategy relying on these forecasts may be vulnerable to extreme price volatility. In Queensland, where high renewable penetration and extreme weather events drive significant price fluctuations, this limitation leads to suboptimal battery scheduling and reduced economic returns [12][13].

### III. UNCERTAINTY DISCOUNTING

#### A. Proposed Methodology

To handle the impact of uncertain price forecasts in this energy arbitrage setting, this study proposes a simple-yet-effective heuristic strategy. The underlying idea is that the farther the forecast is in time from the current instant, i.e., higher the lead time $(n)$, the higher is the uncertainty in it. This stems from a simple and widely applicable intuition: "The usefulness of the information available at instant $k$ in predicting prices at future points $k+i$ and $k+j$ $(i < j)$ will depend on both $i$ and $j$, with the information being more relevant for $k+i$ than for $k+j$."

The simplest way to achieve this is by weighing the forecast prices in a way such that the weights progressively decrease with lead time, $n$, which is the distance of the forecast instance from the current instant. This study proposes three different weighing/discounting mechanisms, inspired from exploration-exploitation trade-off strategies, as shown below:

1. Simulated Annealing:
$$\gamma_n = \exp\left(-\frac{\gamma_0(n-1)}{T_k}\right) \quad\quad - (7)$$

2. Cosine Annealing:
$$\gamma_n = \frac{1}{2} + \frac{1}{2}\cos\left(\frac{(n-1)\pi}{T_k}\right) \quad\quad -(8)$$

3. Power Law:
$$\gamma_n = \gamma_0^{n-1} \quad\quad -(9)$$

where $n \in [1, 2, \dots, T_k]$ and $\gamma_0$ is the initial weight factor. The progression of these weighing factors with $\gamma_0$=0.95 as a function of $n$ is depicted in Figure 2. Three different discounting strategies are proposed mainly to observe the impact of choosing a specific discounting path and rate on the optimization output. The Power Law strategy reduces the weights very fast in time, cosine annealing follows an almost linear path in decreasing the weights while simulated annealing falls somewhere in between the two.

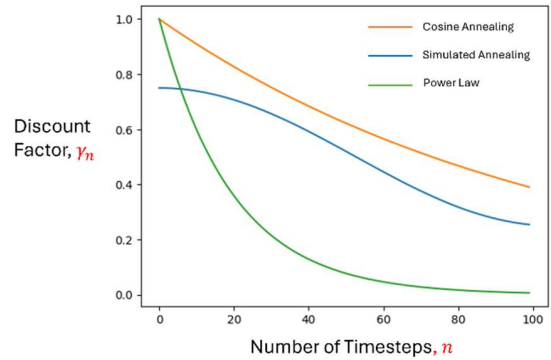

**Figure 2:** Progression of weighing factors as a function of lead time

Incorporating them, the uncertainty-aware MPC is formulated as shown below:
$$\max_{\mathbf{P_k}} \quad \mathbf{c_k^T \Gamma P_k} - \lambda \left\|\mathbf{P_k \Gamma^{-1}}\right\|_s \quad\quad - (10)$$
s.t. (1)-(6)

where $\mathbf{\Gamma} = diag([\gamma_1, \gamma_2, \dots, \gamma_n, \dots, \gamma_{T_k}])$ is the diagonal weighing matrix. The first term in the objective weighs the pre-dispatch price forecasts according to their lead times with lower weights at the tail end of the forecast. This ensures that the weight is higher for the spread components of more certain lead times. The second component is a regularization term with the order of the norm denoted by $s$, that prevents the power outputs of the battery from reaching high values frequently and further decreases the impact of farther forecast lead times. $\lambda$ is the regularization factor.

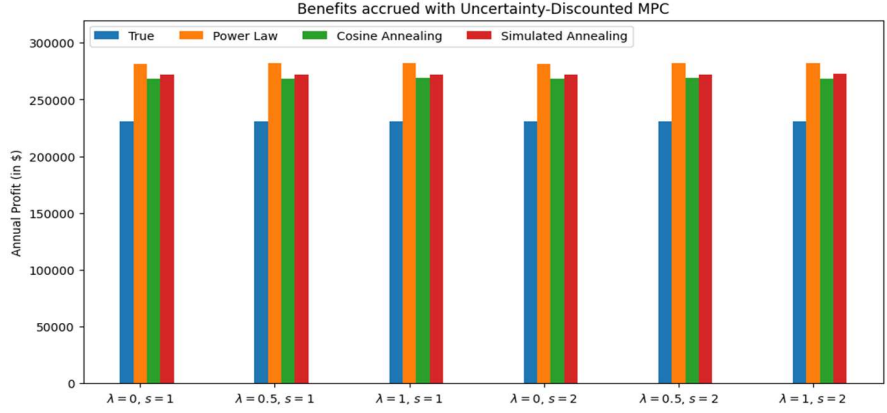

**Figure 3:** Comparison of Annual Profits accrued using different uncertainty discounting strategies with $\gamma_0 = 0.95$. The undiscounted standard MPC results are represented as the True value in blue bars.

**TABLE I:** Annual Benefits (in AUD) accrued from Discounted MPC under different discounting schemes. The maximum values in each column are made **bold**.

| Uncertainty Weighing Scheme | $s$ | $\gamma_0$ | 0.95 | | | 0.99 | | |
|---|---|---|---|---|---|---|---|---|
| | | $\lambda$ | 1 | 0.5 | 0 | 1 | 0.5 | 0 |
| **Simulated Anneal (7)** | 1 | | 272035.1 | 271893.4 | 271712.3 | **273529.7** | **273404.7** | 273334.9 |
| | 2 | | 272590.8 | 272280.5 | 271721.7 | 273283.1 | 273256.5 | **273364.4** |
| **Cosine Annealing (8)** | 1 | | 268877 | 268723.5 | 268543.4 | 268877 | 268723.5 | 268543.4 |
| | 2 | | 268485.4 | 269169.5 | 268582.1 | 268480.8 | 269163.6 | 268578 |
| **Power Law (9)** | 1 | | **282015.6** | **281880.6** | 281179.3 | 256313.2 | 256169.6 | 256109.4 |
| | 2 | | 281990 | 281842.7 | **281189.9** | 256457.8 | 256202.4 | 256129.5 |
| **Standard MPC (1)** | | | 230699.89 | | | | | |

### B. Experimental Details

AEMO's pre-dispatch price forecasts from Nov 2023 – Nov 2024 were considered for this study [14]. Following UQ's operational limits on the Tesla Battery, the upper and lower SOC limits were fixed at 0.1 and 1 respectively. This is to ensure that the battery can be used for FCAS events. $P_l$ and $P_u$ in Eq. (4) were fixed at -1.1 and 1.1 respectively. $E_{nom} = 2.2$ MWh and a battery efficiency of 0.95 is considered. Different instances of the uncertainty-discounted MPC in Eq. (10) were run by varying the hyperparameters, $\gamma_0 \in \{0.99, 0.95\}, \lambda \in \{0, 0.5, 1\}$ and $s \in \{1, 2\}$. The total benefits accrued over the period of one year is calculated as follows:

$$\text{Annual Profit} = \sum_{k=1}^{N} P_k^1 \left( \frac{1}{6} \sum_{j=1}^{6} \mu_{kj} \right)$$

where $\mu_{kj}$ is the true spot price at the $j^{th}$ 5 min interval of the $k^{th}$ 30 min interval, $P_k^1$ is the dispatch value in the first timestep of the plan computed by MPC and $N$ is the total number of timesteps in one year. Price settlements in the National Electricity Market happens in 5 min intervals and the average price over one 30 min interval is multiplied with the optimal power dispatch value to get the actual benefits accrued [12].

## IV. RESULTS

The results of the experiments conducted in Section III-B are shown in Table I and Figure 3. Using the standard undiscounted MPC (1), the battery accrues an arbitrage profit of AU\$ 230699.89 in the period of Nov'23 – Nov'24. Compared to this, the simple heuristic discounting strategies lead to significantly higher profit margins. The maximum increased profits of 22.24% are leveraged by using the Power Law weighing strategy (9) with an initial discount factor of 0.95 regularization coefficient of 1 and norm order 1. The simulated annealing weighing strategy (7) accrues an average 18% increase in arbitrage profits, while cosine annealing (8) increases the profit margin by around 16%. Higher profits are encountered with an initial discount factor of 0.95 compared to that of 0.99, indicating that tail end forecasts need to be weighed exponentially low compared to the initial ones. This validates the heuristic assumption of Section III-A. On other fronts, there are slight variations in profit with varying values of regularization coefficient, $\lambda$, with $\lambda = 1$ being slightly better on average. The L1 regularization norm, i.e, when s = 1, performs slightly better compared to the L2 norm in most cases. This is regardless of the fact that the L1 norm induces sparsity in the optimal power schedule.

One of the key points to note here is that these improvements have been made without requiring any information about the price predictive model or its input-output values. This agnosticism allows the proposed strategy to be universally applicable across predictive model architectures and applications. This also implies that the discounting factors can be pre-computed before optimization and used directly as weighing factors with O(1) complexity, rendering it faster and more effective than any existing approaches leveraging the predictive model in any capacity[15][16]. Further, applying these strategies to the downstream optimization tasks require linear operations in the parameter space. If the original objective function is given by $f(x)$, the uncertainty discounted objective can be represented as $Wf(x) + c$ where $W$ is the weighing factor matrix and $c$ is the regularization sum as shown in (10), reducing it to a simple plug and play operation. These lack of complexities in computation and application to downstream optimization while still being effective in rendering uncertainty awareness makes the proposed heuristic discounting approach an extremely powerful tool.

## V. Conclusion

This study highlights the critical role of uncertainty awareness in predictive energy management, particularly in battery energy arbitrage where market volatility presents significant operational challenges. A case study emulating the University of Queensland's Tesla Battery is performed. The traditional MPC framework used by the UQ battery is designed and its reliance on AEMO's deterministic pre-dispatch price forecasts, which fail to capture the extreme non-stationary behaviors characteristic of high-renewable penetration regions like Queensland is explained. The absence of uncertainty quantification in AEMO's pre-dispatch spot market forecasts limits the effectiveness of existing arbitrage strategies, leaving decision-makers vulnerable to forecast inaccuracies.

By integrating post-hoc heuristic uncertainty discounting techniques, this study demonstrates a more than 20% improvement in arbitrage revenue, offering a practical solution that enhances decision robustness without requiring predictive model access. The proposed strategies are computationally efficient, require negligible changes in the existing predictive control structures and are agnostic of the predictive model. These findings underscore the necessity of uncertainty-aware control mechanisms in energy markets and lay the groundwork for future research in adaptive optimization strategies, ensuring resilient and economically viable energy management in volatile market conditions.

## Acknowledgment

The author gratefully acknowledges the contributions of Prof. Tapan K. Saha, Professor, The University of Queensland for his invaluable feedback on this work.